\documentclass[twocolumn,pre]{revtex4}

\usepackage{graphicx,setspace}
\usepackage{dcolumn}
\usepackage{bm}
\begin{document}

\title{Coherence resonance in an unijunction transistor relaxation oscillator}%
\author{Md. Nurujjaman}
\email{jaman_nonlinear@yahoo.co.in}
\affiliation{Plasma Physics Division, Saha Institute of Nuclear Physics,
1/AF, Bidhannagar, Kolkata -700064, India.}

\author{P.S. Bhattacharya}
\affiliation{Plasma Physics Division, Saha Institute of Nuclear Physics,
1/AF, Bidhannagar, Kolkata -700064, India.}
\author{Sandip Sarkar}
\affiliation{Microelectronics Division, Saha Institute of Nuclear Physics,
1/AF, Bidhannagar, Kolkata -700064, India.}
\author{A.N. Sekar Iyengar}
\email{ansekar.iyengar@saha.ac.in}
\affiliation{Plasma Physics Division, Saha Institute of Nuclear Physics,
1/AF, Bidhannagar, Kolkata -700064, India.}

\begin{abstract}
The phenomenon of coherence resonance (CR) is investigated in an unijunction transistor
relaxation oscillator (UJT-RO) and quantified by estimating the normal variance (NV).
Depending upon the measuring points two types of NV curves have been obtained. We have
observed that the degradations in coherency  at higher noise amplitudes in our system is
probably the result of direct interference of coherent oscillations and the stochastic
perturbation. Degradation of coherency may be minimal if this direct interference of noise
and coherent oscillations is eliminated.

\end{abstract}

\maketitle
The study of  the constructive role of noise, which is generally termed as  coherence resonance (CR),
in  threshold or excitable systems, has received considerable attention in the last twenty years and
has been observed in many physical, chemical and
biological and electronics systems
 ~\cite{jstatphys:sigeti,prl:gang,prl:wiesenfeld,pre:strogatz,prl:pikovsky,prl:locher,pre:bascones,pre:alonso,rmp:sagues,pre:dubbeldam,prl:stock,fnl:stock,physrep:linder,pre:jaman,pre:santos,pre:santos1,pre:lindner,prl:Giacomelli,pre:lee,pre:miyakawa}. The basic characteristic of a threshold system is that it shows fixed point and limit cycle attractors below and above a threshold respectively. When the system resides at a fixed point, it reacts to external perturbations in two different ways depending upon the perturbation amplitude. When the amplitude is small to cross the threshold, the system remains at a stable state, but  if it is large enough to cross,  a large amplitude excursion is produced in the output through limit cycle oscillations, before settling back to the fixed point. Application of the time dependent external stochastic perturbation  produces coherent limit cycle oscillations, depending upon the amplitude of the perturbation and maximum coherency has been observed for an optimum noise and coherency decreases for higher noise amplitudes \cite{prl:pikovsky,pre:jaman,pre:santos,pre:santos1,pre:lindner,prl:Giacomelli,pre:lee,pre:miyakawa,pla:Sethia}. The decrease in the coherency with increase in noise intensity was explained as due to the increase in variation of the excursion time of the limit cycles for small threshold and large excursion time where increase in mean excursion time is negligible~\cite{prl:pikovsky}. But if stochastic perturbation helps the system only to cross the threshold,  then at higher amplitudes, the system will remain at an excitable state~\cite{phd:Ullner}  producing only limit cycle oscillations which has maximum coherency, that contradicts the above explanation.

In this letter we have presented experimental results that is consistent with latter explanation
by means of an UJT-RO which exhibits relaxation oscillations over a certain range of emitter
voltages and is used extensively in triggering circuits~\cite{book:mottershed,pra:Koepke}. We
also show here that the destruction of coherency is due to direct interference of noise on the
limit cycle oscillations which is consistent with experimental results presented
in Ref.~\cite{prl:Giacomelli}.

\label{section:setup}

\begin{figure}[ht]
\centering
\includegraphics[width=8.5cm]{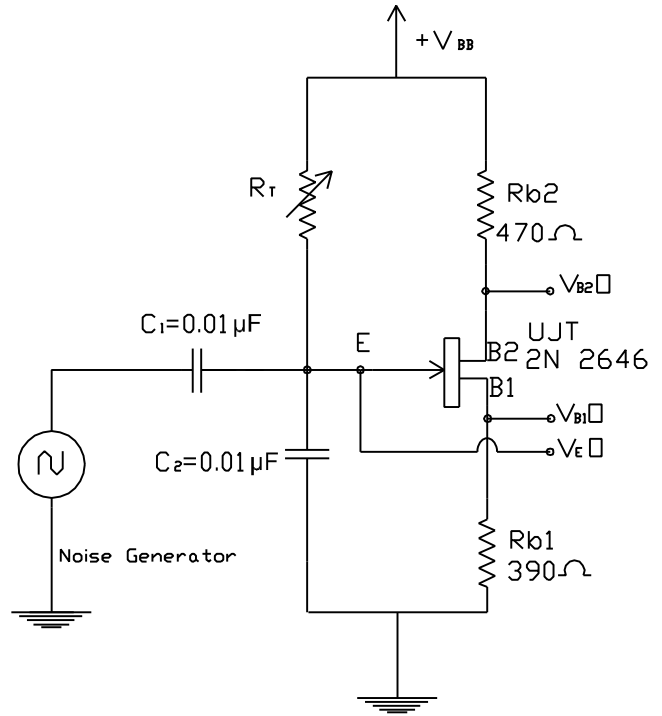}
\caption{Circuit Description of the UJT Relaxation Oscillator}
\label{fig:ckt}
\end{figure}

\begin{figure*}[ht]
\centering
\includegraphics{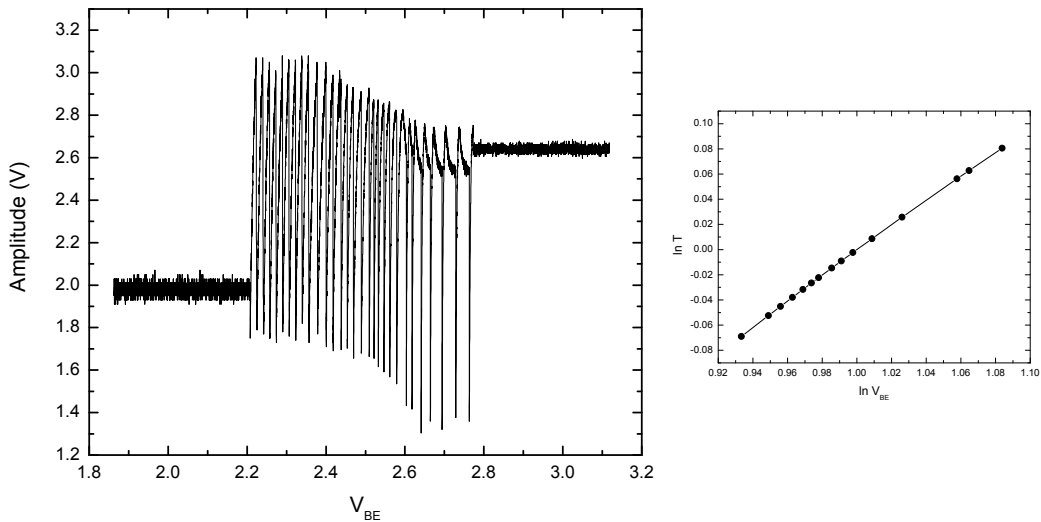}
\caption{Left panel: Experimental bifurcation diagram constructed by increasing the
voltage $V_{E}$ at the output terminal $V_EO$. It shows that the parameters are divided
into two regions: the fixed point region where there are no oscillations in the output
and oscillatory region where oscillations are observed. Right panel: $\ln T$ vs $\ln V_E$
curve fitted by a straight line indicates a power law behavior of the increment of time period (T). }
\label{fig:bifurcation}
\end{figure*}

The experiment has been carried out in an UJT-RO whose schematic circuit diagram has
been shown in Figure~\ref{fig:ckt}.  For the present experiments, the oscillator was
operated at $V_{BB}= 4.05$ V. Depending on the voltage ($V_E$) between emitter (E) and
base 1 (B1) periodic oscillations were observed and the outputs were recorded at two points
E (V$_E$O) and B1 (V$_{B1}$O) respectively. $V_E$ which is the control parameter in the
present experiment, can be varied by using a variable resistance $R_T$. A noise source
was coupled at point E through the capacitor $C_1$  for noise induced experiments.

Before studying the noise-invoked resonance dynamics, we characterized the autonomous
behavior of the oscillator. Depending upon the control parameter ($V_{E}$) the relaxation
oscillations were observed for a particular range of $V_{E}$. Fig~\ref{fig:bifurcation} (left panel)
shows the experimental bifurcation diagram using $V_{E}$ as the control parameter, and is seen that
around $V_{E}\approx2.5$ V, limit cycle oscillations were observed and vanish at about $V_E\approx3 V$.
The time period of these oscillations increases with $V_{E}$. The graph of $\ln T$ vs $\ln V_{E}$
(right panel)  follows a power law behavior with an exponent $\approx0.99$. So $V_{E}\approx3 V$
defines the threshold of the system for the particular choice of parameters.

To summarize, at $V_{E}$ values below 2.5 V and above 3 V excitable fixed point response was
found.  For the purpose of the present experiments the set point was kept above the
threshold  (V$_E\approx$ (3 V) and noise induced coherent responses were recorded
at V$_E$O and V$_{B1}$O respectively.


\begin{figure*}[ht]
\centering
\includegraphics[width=12cm]{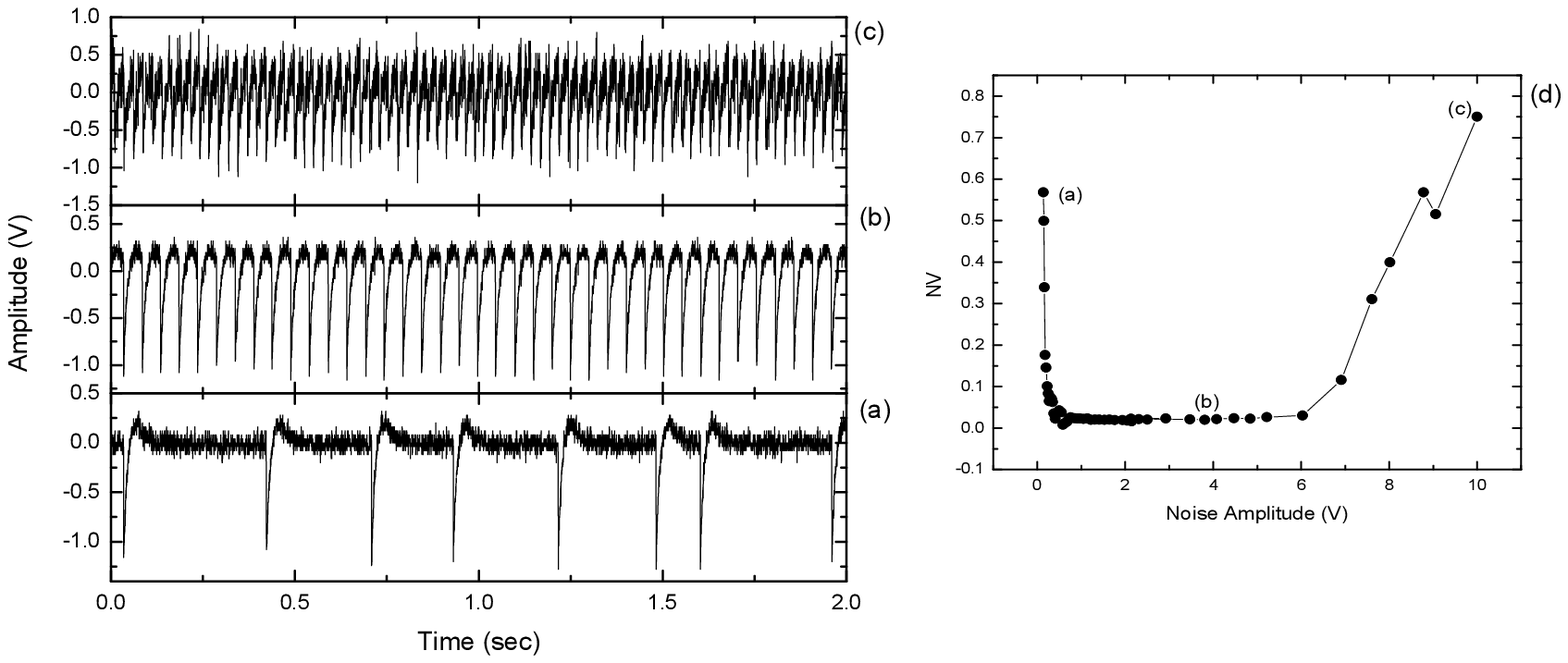}
\caption{Emergence of coherence resonance for the output recorded
at emitter (V$_E$O): The right panel shows the NV as a function of
noise amplitude for the experiments performed at $V_{E}=3.142 V$.
Left panel: The time series of the output for (a) low (0.14 V),
(b) optimum (3.81 V) and (c) high-level (10 V) noise.  }
\label{fig:nv2}
\end{figure*}

 \begin{figure*}[ht]
\centering
\includegraphics[width=12cm]{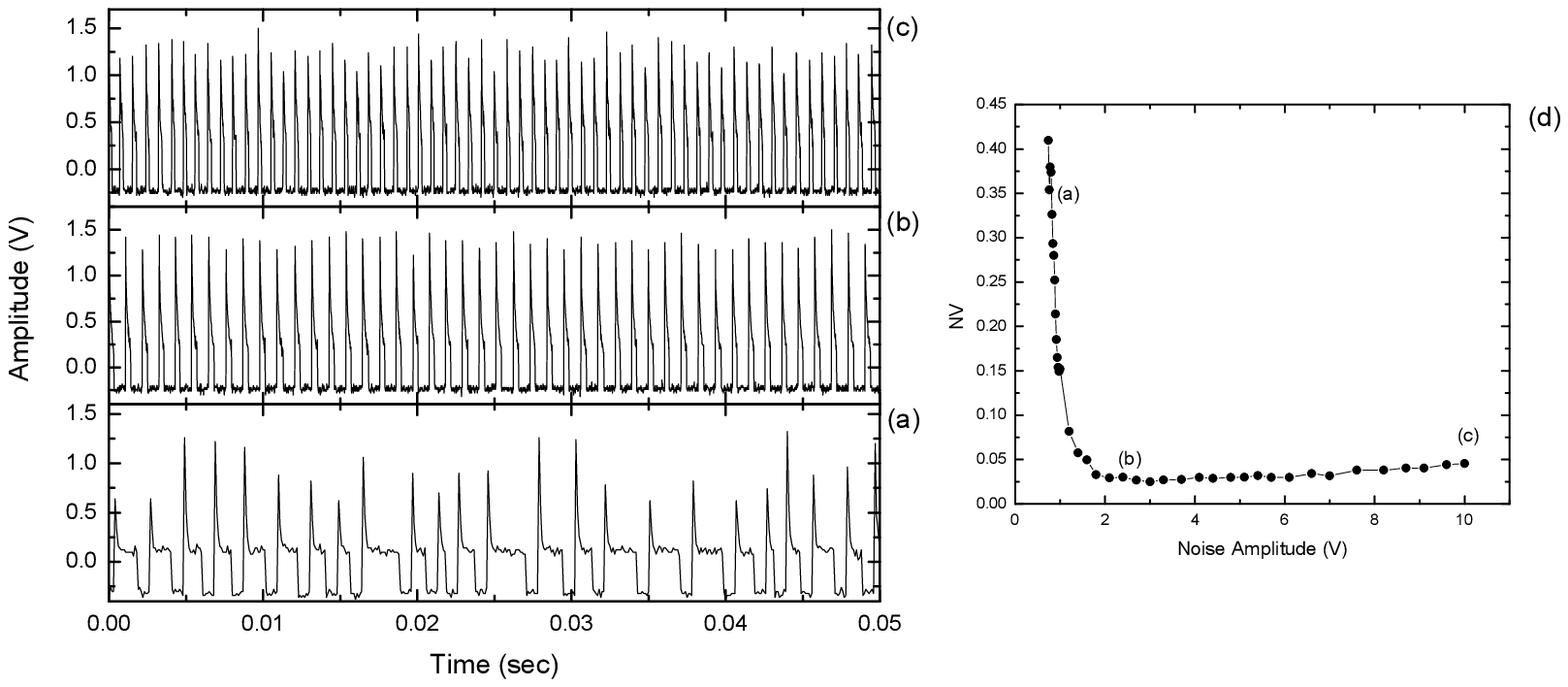}
\caption{Emergence of coherence resonance for the output recorded at base1 (V$_{B1}$O):
The right panel shows the NV as a function of noise amplitude for the experiments
performed at $V_{E}=3.142 V$. Left panel: The time series of the output for (a) low (0.76 V),
(b) optimum (2.4 V) and (c) high-level (10 V) noise.  }
\label{fig:NVghraph}
\end{figure*}

For the experiments on coherence resonance, the resistance $R_T$ was chosen such that
the output of the UJT-RO exhibits fixed point behavior in the absence of noise and a
set point  was chosen far from the threshold so that the system always remains in a stable
state under the influence of the intrinsic noise and parametric drifts. A Gaussian noise
was superimposed on the bias voltage through the capacitor $C_1$ [Fig~\ref{fig:ckt}] and
the regularity of the provoked dynamics which depends upon noise intensity, were analyzed.
The normalized variance (NV) was used to quantify the induced coherency. It is defined
as  $NV=std(t_p)/mean(t_p)$, where, $t_p$ is the time taken between successive peaks.
It is evident that the more the coherent the induced oscillations, the lower the value
of the computed NV~\cite{pre:jaman}. For pure oscillations, NV tends to zero.

Figs.~\ref{fig:nv2}(a)$-$\ref{fig:nv2}(c) (left panel) show the time series of the
output recorded at V$_{E}$O for different noise levels and Fig.~\ref{fig:nv2}(d) is
the experimental NV curve as a function of noise amplitude. The point (a) in Fig.~\ref{fig:nv2}
(d) corresponds to the time series shown in Fig.~\ref{fig:nv2}(a) and is associated with a
low level of noise where the activation threshold is seldom crossed, generating a sparsely
populated irregular spike sequence. As the noise amplitude is increased, the NV decreases
and reaches a minimum.  Fig.~\ref{fig:nv2}(b) shows the time series corresponds to a the
point (b) in Fig.~\ref{fig:nv2}(d) for maximum regularity at an optimum noise level.
At higher amplitudes of the superimposed noise, the observed regularity is destroyed
manifested by an increase in the NV; label (c) in Fig.~\ref{fig:nv2}(d), and the corresponding
time series has been shown in Fig.~\ref{fig:nv2}(c). This is a consequence of the dynamics being
dominated by noise.

Fig~\ref{fig:nv2}(d) shows that the system attains a coherent state with increase in noise
amplitude and stays at that state for a wide range of amplitudes before being degraded at
higher values. Initially, the inter-peak time is dominated by the activation time of the
limit cycle oscillations which varies significantly at low noise levels, and with increase
in the noise amplitudes the activation time decreases resulting in a rapid fall
of NV~\cite{prl:pikovsky,pre:miyakawa}. When the noise amplitude is such that it crosses
the threshold very often, i.e., number of crossing is much larger than the excursion frequency
of the oscillations,  it remains in the excited state, resulting in coherent oscillations
leading to a flat minimum in the NV curve [Fig~\ref{fig:nv2}(d)]. Direct interference of the
strong noise distorts the limit cycle oscillations particularly, the peaks and since the
excursion time is almost independent of noise amplitude~\cite{prl:pikovsky,pre:miyakawa}
the increase in NV at higher noise is probably due to  the increase in the variation of
the inter-peak distances of the distorted peaks of the limit cycle oscillations~\cite{prl:Giacomelli}.

Figs~\ref{fig:NVghraph}(a)$-$\ref{fig:NVghraph}(c)  show the time series of the output
recorded at V$_{B1}$O of the oscillator  for different noise levels and Fig.~\ref{fig:NVghraph}
(d) is the experimental NV curve as a function of noise amplitude. As noise is increased from
its minimum value, the NV rapidly reaches a minimum  and remains constant and shows a slight
tendency to increase at higher noise amplitudes. Fig~\ref{fig:NVghraph}(a) shows the time
series of irregular spiking at low noise  levels and its corresponding NV point is
(a) in Fig~\ref{fig:NVghraph}(d). Figs~\ref{fig:NVghraph}(b) and \ref{fig:NVghraph}
(c) show the time series for two NV points (b) and (c) in Fig~\ref{fig:NVghraph}(d)
at moderate and maximum noise respectively. It is seen that increasing the noise amplitude
does not result in significant difference in the measure of the coherency, i.e., in NV.
Fig~\ref{fig:NVghraph}(c)  shows that there is no significant distortion in the limit
cycle at higher intensity of noise though it is prominent in the first case [Fig~\ref{fig:nv2}(c)]
and this is probably because, noise is blocked by some internal self organization of the UJT which,
prevents the destructive effect of  the stochastic perturbation  on the oscillations. In this case
the higher amplitude noise only invokes permanent excitation in the system, resulting in coherent
oscillations, and hence small NV. This also shows the robustness of the UJT against the destructive
effect of noise in triggering circuits.

These observations indicate that the destructive effect of large stochastic perturbation
probably depends upon  the system properties, relevant parameters, etc., and  the system
which can block their direct interference, may not exhibit significant increase in NV at
higher amplitude.
\begin{figure}[ht]
\centering
\includegraphics[width=3.5in]{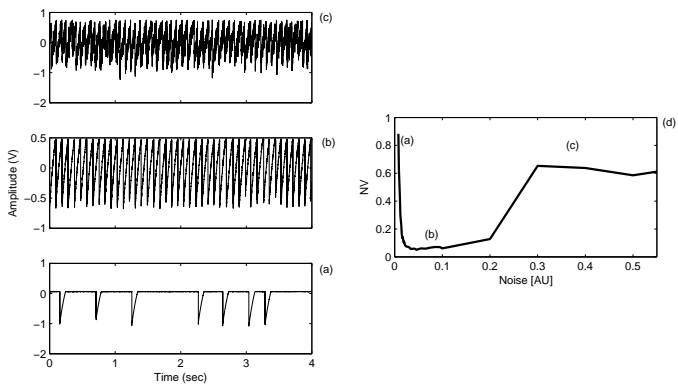}
\caption{Emergence of coherence resonance in the pspice simulation for the output
recorded (V$_{E}$): The right panel shows the NV as a function of noise amplitude
for the experiments performed at $V_{E}=2.75 V$. Left panel: The time series of the
output for (a) low , (b) optimum and (c) high-level noise.}
\label{fig:sim1}
\end{figure}

\begin{figure}[ht]
\centering
\includegraphics[width=3.5in]{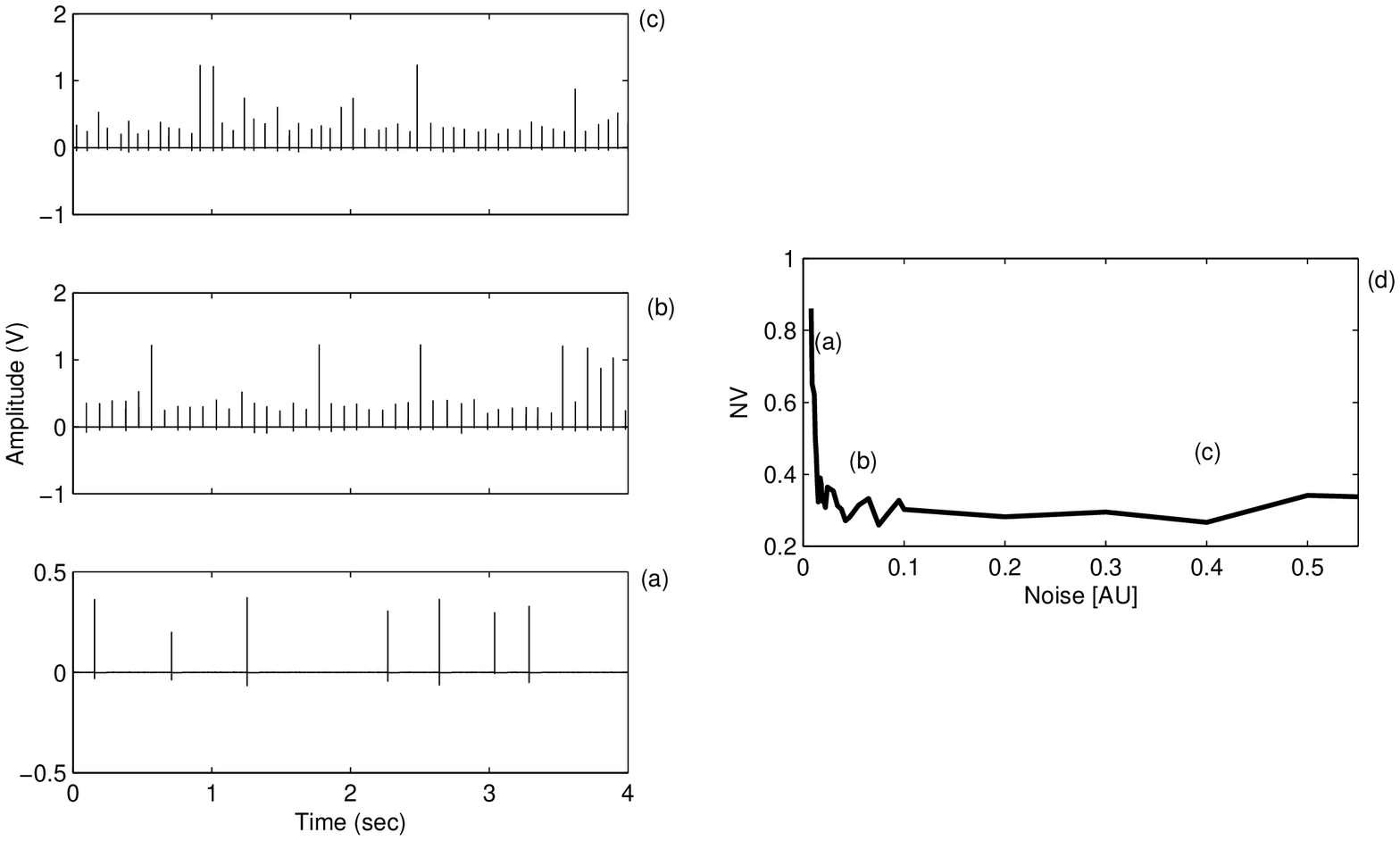}
\caption{Emergence of coherence resonance for the output recorded at base1 (V$_{B1}$O):
The right panel shows the NV as a function of noise amplitude for the experiments
performed at $V_{E}=2.75 V$. Left panel: The time series of the output for (a) low ,
(b) optimum and (c) high-level noise.}
\label{fig:sim2}
\end{figure}

To validate our experimental results, PSPICE   simulation of the
UJT-RO were carried out  using  same values for the capacitors and
resitors as  in the real experiments. The only exception in the
simulation was that $R_T$ was chosen slightly higher than the
experimental value to obtain relaxation oscillations . The
relaxation  oscillations were observed around $V_{E}\approx 2.6$ V
and ceased around $\approx 2.65 V$. For noise induced experiments,
$V_{E}$ was set at 2.75 V and noise generated using MATLAB was fed
into the circiut and measurement were performed at two points E
and B1 respectively. Fig~\ref{fig:sim1}(d) shows the  PSPICE
simulated NV curve and Figs~\ref{fig:sim1}(a)-(c) the three time
series corresponding to the points (a)-(c) shown on the same NV
plot [Fig~\ref{fig:sim1}(d)] for the data recorded at $V_E$. The
NV curve and the time series for the simulated results show almost
similar  features of the experimental measurements shown in
Fig~\ref{fig:nv2}. Similarly, Fig~\ref{fig:sim2}(d) shows the
simulated NV curve for the data recorded at $V_{B1}$ and
Figs~\ref{fig:sim2}(a)-(c) the three time series corresponding to
the points (a)-(c) shown on the same NV plot
[Fig~\ref{fig:sim2}(d)]. Both PSPICE simulated NV curves
[\ref{fig:sim1}(d) and \ref{fig:sim2}(d)] shows a behavior very
similar to the  experimental NV curves [Fig~\ref{fig:nv2}(d) and
Fig~\ref{fig:NVghraph}(d)].


In conclusion, the effect of noise has been studied experimentally
in UJT-RO, demonstrating the emergence of CR via purely stochastic
fluctuations. PSPICE  simulation also shows similar kind of
behavior. We have also shown that depending upon system
properties, coherency may or may not be destroyed at high
amplitude noise. Both kinds of NV may be observed in same system
depending upon the relevant parameters.

We would like to acknowledge the help of Mr. Dipankar Das and the other members of the Plasma
Physics Division and Microelectronics Division during the experiments. One of the authors (ANSI)
would like to thank Prof. P. Parmananda for some useful discussion on noise invoked dynamics.


\begin{thebibliography}{100.}
\bibitem{jstatphys:sigeti} {David Sigeti and Werner Horsthemke, J. Stat. Phys. \textbf{54}, 1217 (1989)}.
\bibitem{prl:gang} {Hu Gang, T. Ditzinger, C. Z. Ning, and H. Haken, Phys. Rev. Lett. \textbf{71},
807 (1993)}.
\bibitem{prl:wiesenfeld} {Kurt Wiesenfeld, David Pierson, Eleni Pantazelou, Cris Dames,
and Frank Moss, Phys. Rev. Lett. \textbf{72}, 2125 (1994)}.
\bibitem{pre:strogatz} {W. J. Rappel and Steven Strogatz, Phys. Rev. E \textbf{50}, 3249 (1994)}.
\bibitem{prl:pikovsky} {Arkady S. Pikovsky and J\"{u}rgen Kurths, Phys. Rev. Lett. \textbf{78},
775 (1997)}.
\bibitem{prl:locher} {M. L\"ocher,  D. Cigna,   and E. R. Hunt, Phys. Rev. Lett. \textbf{80}, 5212 (1998)}.
\bibitem{pre:bascones} {R. B\'ascones,  J. Garc\'ia-Ojalvo,   and J. M. Sancho,
Phys. Rev. E \textbf{65}, 061108 (2002)}.
\bibitem{pre:alonso} {S. Alonso,  I. Sendi\~na-Nadal,    V. P\'erez-Mu\~nuzuri,
J. M. Sancho,  and F. Sagu\'es,   Phys. Rev. lett. \textbf{87}, 078302 (2001)}.
\bibitem{rmp:sagues} {F. Sagu\'{e}s, J. M. Sancho, J. Garc\'ia-Ojalvo,
Rev. Mod. Phys. \textbf{79}, 829 (2007)}.
\bibitem{pre:dubbeldam} {Johan L. A. Dubbeldam, Bernd Krauskopf, and Daan Lenstra,
Phys. Rev. E \textbf{60}, 6580 (1999)}.
\bibitem{prl:stock} {N. G. Stocks, Phys. Rev. Lett. \textbf{84}, 2310 (2000).}
\bibitem{fnl:stock} {N. G. Stocks, D. Allingham, and R. P. Morse, Fluctuation and
Noise Letters \textbf{2}, L169 (2002).}
\bibitem{physrep:linder} { Lindner B, Garcia-Ojalvo J., Neiman A. and Schimansky-Geier L,
Phy. Rep.  \textbf{392}, 321 (2004).}
\bibitem{pre:jaman} {Md. Nurujjaman, A.N.Sekar Iyengar, and P. Paramanda
Phy. Rev. E \textbf{78}, 026406 (2008).}
\bibitem{pre:santos} { Gerando J. Escalera Santos, and P. Paramanda,
Phy. Rev.E \textbf{65}, 067203 (2002).}
\bibitem{pre:santos1} { Gerando J. Escalera Santos, M. Rivera, M. Eiswirth, and P. Paramanda,
Phy. Rev.E \textbf{70}, 021103 (2004).}
\bibitem{pre:lindner} { Benjamin Lindner and Lutz Schimansky-Geier, Phy. Rev.E \textbf{61}, 6103 (2000).}
\bibitem{prl:Giacomelli} {Giovanni Giacomelli, Massimo Giudici, Salvador Balle,
and Jorge R. Tredicce, Phys. Rev. Lett. \textbf{84}, 3298 (2000)}.
\bibitem{pre:lee} { Sang-Gui Lee, Alexander Neiman, and Seunghwan Kim, Phy. Rev.E \textbf{57}, 3292 (1998).}
\bibitem{pre:miyakawa} { Kenji Miyakawa and Hironobu Isikawa, Phy. Rev.E \textbf{66}, 046204 (2002).}
\bibitem{pla:Sethia} {Gautam C. Sethia, J\"{u}rgen Kurths, and Abhijit Sen, Phys. Lett. A \textbf{364}, 227 (2007). }
\bibitem{phd:Ullner} {Ekkehard Ullner, PhD thesis, \emph{Noise-induced phenomena of signal
transmission in excitable neural models} (Chapter 1).}
\bibitem{book:mottershed} {Mottershed Allen:  \emph{Electronic Devices And Circuits},
Prentice Hall of India,1980, 538-540. }
\bibitem{pra:Koepke} {M.E. Koepke and D.M. Hartley, Phy. Rev.A \textbf{44}, 6677 (1991).}
\end{thebibliography}
\end{document}